\documentclass{sig-alternate}

\usepackage{url}
\usepackage{subfigure}

\usepackage{rc-book/rc-listing-2.14}
\usepackage{rc-book/rc-math-2.14}
\usepackage{rc-book/rc-tabular-2.14}

\input{macros/Dirty-Tricks.tex}

\begin{document}

\title{Anatomy of Graph Matching based on an XQuery and RDF Implementation}

\numberofauthors{2}

\author{
\alignauthor
Carlos R. Rivero\\
    \affaddr{Department of Computer Science}\\
    \affaddr{University of Idaho, USA}\\
    \email{crivero@uidaho.edu}
\alignauthor
Hasan M. Jamil\\
    \affaddr{Department of Computer Science}\\
    \affaddr{University of Idaho, USA}\\
    \email{jamil@uidaho.edu}
}

\maketitle
\begin{abstract}
Graphs are becoming one of the most popular data modeling paradigms since they are able to model complex relationships that cannot be easily captured using traditional data models. One of the major tasks of graph management is graph matching, which aims to find all of the subgraphs in a data graph that match a query graph. In the literature, proposals in this context are classified into two different categories: graph-at-a-time, which process the whole query graph at the same time, and vertex-at-a-time, which process a single vertex of the query graph at the same time. In this paper, we propose a new vertex-at-a-time proposal that is based on graphlets, each of which comprises a vertex of a graph, all of the immediate neighbors of that vertex, and all of the edges that relate those neighbors. Furthermore, we also use the concept of minimum hub covers, each of which comprises a subset of vertices in the query graph that account for all of the edges in that graph. We present the algorithms of our proposal and describe an implementation based on XQuery and RDF. Our evaluation results show that our proposal is appealing to perform graph matching.
\end{abstract}




\section{Introduction}

Nowadays, there is an increasing interest in using graphs to represent data~\cite{journals/prl/WeberLD12}. This is due to the fact that graphs allow to model complex relationships that cannot be easily captured using traditional data models like the relational model~\cite{conf/sigmod/HeS08}. One of the crucial tasks of graph management is graph matching, which aims to find all of the subgraphs in a data graph that match a query graph~\cite{journals/jiis/BhattacharjeeJ12}.

In the literature, proposals to perform graph matching can be broadly classified into two different categories: graph-at-a-time and vertex-at-a-time. In the former, the whole query graph is processed at the same time, in which GraphQL~\cite{conf/sigmod/HeS08}, QuickSI~\cite{journals/pvldb/ShangZLY08}, Ullman~\cite{journals/jacm/Ullmann76} and VFLib~\cite{conf/iciap/CordellaFSV99} are the most representative proposals. In the latter, a single vertex of the query graph is processed at the same time, in which SAPPER~\cite{journals/pvldb/ZhangYJ10} and TALE~\cite{conf/icde/TianP08} are the most representative proposals. The most appealing advantage of the vertex-at-a-time proposals is that they are able to prune the search space using matching conditions, such as vertex indices~\cite{journals/pvldb/ZhangYJ10}, path indices~\cite{conf/icpr/GiugnoS02}, or frequent structures~\cite{conf/edbt/ZhangLY09}.

Our goal is to devise a new vertex-at-a-time proposal that is based on graphlets, each of which is a three tuple that comprises a vertex of a graph, all of the immediate neighbors of that vertex, and all of the edges that relate those neighbors. In our proposal, both the data and query graphs are represented as sets of graphlets. To process the query graph over the data graph, we first reduce the number of graphlets in the query graph based on a new notion of edge covering in graph theory, called the minimum hub cover, which comprises a subset of vertices in a graph that account for all of the edges in that graph. Note that there may exist more than one minimum hub covers for the same query graph. Then, our proposal sort the vertices of the minimum hub cover to reduce the search space as much as possible. Finally, we use a number of algorithms to compute the subgraphs of the data graph that match the query graph.

In this paper, we assume that the minimum hub covers for a given query graph are already computed. In Section~\ref{sec:preliminaries}, we introduce the main concepts that we use throughout the paper. We present a discussion on how to select the best minimum hub cover and the best ordering of this minimum hub cover in Section~\ref{sec:mhc}. Section~\ref{sec:graph-matching} describes the algorithms to match a query graph over a data graph. In Section~\ref{sec:implementation}, we describe an implementation of the previous algorithms using XQuery and RDF. Section~\ref{sec:evaluation} presents some initial results that we have obtained using this implementation. Finally, Section~\ref{sec:conclusions} recaps on our main conclusions. 
\section{Preliminaries}
\label{sec:preliminaries}

\begin{figure}
    \subfigure[Data graph.]{
        \includegraphics[scale=0.75]{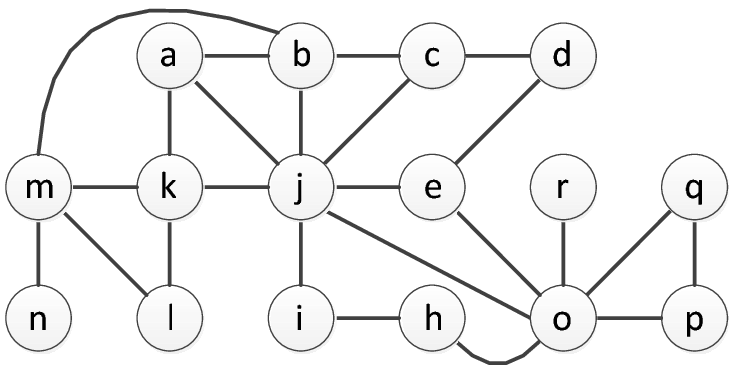}
    }
    \quad
    \subfigure[Query graph.]{
        \includegraphics[scale=0.75]{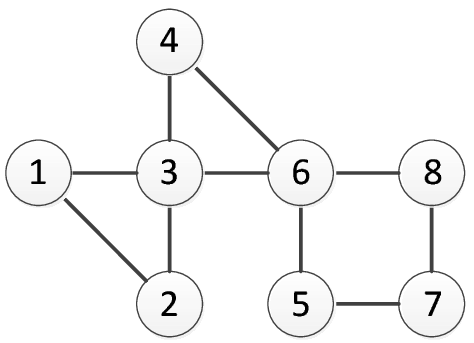}
    }
\caption{Motivating example.}
\label{fig:motivating-example}
\end{figure}

To illustrate our proposal, we use the data and query graphs that are shown in Figure~\ref{fig:motivating-example}. The data graph comprises a total number of 16 vertices and 24 edges connecting those vertices. The graphlets that result of these data graph are the following:

\begin{formulae}[l l l]
 {<}a, & \{b, j, k\}, & \{\{b, j\}, \{k, j\}\}{>} \\
 {<}b, & \{a, c, j, m\}, & \{\{a, j\}, \{c, j\}\}{>} \\
 {<}c, & \{b, d, j\}, & \{\{b, j\}\}{>} \\
 {<}d, & \{c, e\}, & \{\}{>} \\
 {<}e, & \{d, j, o\}, & \{\{j, o\}\}{>} \\
 {<}h, & \{i, o\}, & \{\}{>} \\
 {<}i, & \{h, j\}, & \{\}{>} \\
 {<}j, & \{a, b, c, e, k, i, o\}, & \{\{a, b\}, \{a, k\}, \{b, c\}, \{e, o\}\}{>} \\
 {<}k, & \{a, j, l, m\}, & \{\{a, j\}, \{l, m\}\}{>} \\
 {<}l, & \{k, m\}, & \{\{k, m\}\}{>} \\
 {<}m, & \{b, k, l, n\}, & \{\{k, l\}\}{>} \\
 {<}n, & \{m\}, & \{\}{>} \\
 {<}o, & \{e, h, j, p, q, r\}, & \{\{e, j\}, \{p, q\}\}{>} \\
 {<}p, & \{o, q\}, & \{\{o, q\}\}{>} \\
 {<}q, & \{o, p\}, & \{\{o, p\}\}{>} \\
 {<}r, & \{o\}, & \{\}{>}
\end{formulae}

Each graphlet is a three tuple in which the first element is the main vertex $v$, the second element is the set of neighbors of $v$, i.e., those vertices that are related to $v$ by an edge, and the third element is the set of boundaries, each of which is a pair of vertices that belong to the neighbors of $v$ and there exists an edge between them. Therefore, we represent a graphlet as the following tuple: ${<}v, N, B{>}$, in which $N$ is the set of neighbors and $B$ is the set of boundaries. A graph is represented as a set of graphlets, and we use the following auxiliary functions to have access to all of the vertices and edges of a given graph $g$: $vertices(g)$ and $edges(g)$, respectively.

The query graph in Figure~\ref{fig:motivating-example} comprises a total number of 8 vertices and 10 edges, whose graphlets are the following:

\begin{formulae}[l l l]
 {<}1, & \{2, 3\}, & \{\{2, 3\}\}{>} \\
 {<}2, & \{1, 3\}, & \{\{1, 3\}\}{>} \\
 {<}3, & \{1, 2, 4, 6\}, & \{\{1, 2\}, \{4, 6\}\}{>} \\
 {<}4, & \{3, 6\}, & \{\{3, 6\}\}{>} \\
 {<}5, & \{6, 7\}, & \{\}{>} \\
 {<}6, & \{3, 4, 5, 8\}, & \{\{3, 4\}\}{>} \\
 {<}7, & \{5, 8\}, & \{\}{>} \\
 {<}8, & \{6, 7\}, & \{\}{>}
\end{formulae}

Our goal is to find all of the subgraphs in the data graph that match the query graph. The final result is a set of solutions, each of which comprises a set of pairs in which the first element, the key, is a vertex in the query graph, and the second element, the value, is a vertex in the data graph. Therefore, we represent a solution $s$ of a query graph $q$ over a data graph $g$ as a set of pairs $(u, v)$, in which $u \in vertices(q)$ and $v \in vertices(g)$. We also define two auxiliary functions: $keys$ and $values$; the former retrieves all of the vertices of the query graph in the solution, and the latter retrieves all of the vertices of the data graph in the solution. Furthermore, every solution has to comprise all of the vertices in the query graph and a subset of vertices of the data graph, i.e., $keys(s) = vertices(q)$ and $values(g) \subseteq vertices(g)$. A sample solution in our motivating example is the following: $\{(1, m), (2, l), (3, k), (4, a), (5, c), (6, j), (7, d), (8, e)\}$.

Our proposal relies on using minimum hub covers, each of which contains a subset of vertices in the query graph whose graphlets cover all of the edges of the original query graph. This entails that processing a minimum hub cover is exactly the same as processing all of the graphlets of the query graph. Therefore, $M \subseteq vertices(q)$ is a minimum hub cover of $q$ if $M$ is the smallest set, and these two conditions hold:

\begin{itemize}
    \item There exists an edge ${<}u, v{>} \in edges(g)$, such that $u \in M$ or $v \in M$.
    \item There exist two edges ${<}u, v{>} \in edges(g)$ and ${<}v, x{>} \in edges(g)$, such that $x \in M$.
\end{itemize}

Note that there may exist more than one minimum hub cover for the same query graph. For instance, in our motivating example, there exist four minimum hub covers: $\{1, 6, 7\}$, $\{2, 6, 7\}$, $\{3, 6, 7\}$, and $\{3, 5, 8\}$. Now, the challenge is to select the minimum hub cover and the ordering that will produce the minimum number of solutions to explore and, therefore, the minimum processing time.
\section{On selecting and ordering minimum hub covers}
\label{sec:mhc}

The main challenge of using minimum hub covers is to select the best one among all possible minimum hub covers for a given query graph, and the best ordering to be processed, so we explore the minimum number of solutions. To perform this, we focus on three different aspects that must be taken into account: 1) The connection of the vertices in the minimum hub cover; 2) The selectivity of the vertices; and 3) The use of the data graph. In the rest of this section we discuss about these three aspects.

Regarding the connection of vertices, it is mandatory to take the connection of the vertices into account when selecting and ordering the minimum hub covers. This entails that, if vertices $u$ and $v$ appear consecutively in the ordering of a minimum hub cover, both graphlets have to be connected by, at least, one vertex. This avoids performing cartesian products when exploring the solutions to the query graph. For instance, in our motivating example, for the minimum hub cover $\{1, 6, 7\}$, we must avoid orderings $(1, 7, 6)$ and $(7, 1, 6)$ since graphlets $1$ and $7$ do not have any vertices in common. The rest of the orderings are suitable candidates since graphlets $1$ and $6$ have vertex $3$ in common, and graphlets $6$ and $7$ have vertices $5$ and $8$ in common.

Regarding the selectivity of the vertices, it is mandatory to ensure the maximum selectivity of the vertices, i.e., we must select first those graphlets that filter the maximum number of graphlets in the data graph. To compute this, we rely on the concept of free neighbors, which are those neighbors that do not appear in the boundaries of a data graphlet. One heuristic is that the most selective graphlet is the one that has more boundaries and less free neighbors. Furthermore, if the number of boundaries is equal to zero, another heuristic is that the most selective graphlet is the one that has more free neighbors.

\begin{table}
\centering
\begin{rctabular}{|r|r|r|}{Graphlet & Boundaries & Free neighbors}
1 & 1 & 0 \\
2 & 1 & 0 \\
3 & 2 & 0 \\
4 & 1 & 0 \\
5 & 1 & 0 \\
6 & 1 & 2 \\
7 & 0 & 2 \\
8 & 0 & 2
\end{rctabular}
\caption{Boundaries and free neighbors of our motivating example.}
\label{tab:graphlet-boundaries-free}
\end{table}

In our motivating example, the number of boundaries and free neighbors of the graphlets for the query graph are shown in Table~\ref{tab:graphlet-boundaries-free}. Using the previous heuristics, we have to select the most selective graphlet among the following: 1, 2, 3, 5, 6, 7, and 8. In this case, 3 is the most selective graphlet since the number of boundaries is two, which is the maximum number of boundaries. We have already discarded minimum hub covers $\{1, 6, 7\}$ and $\{2, 6, 7\}$, since they do not contain graphlet 3. Then, in the following step, we have to select the most selective graphlet among the following: 5, 6, 7, and 8; the most selective one is graphlet 5, which has more boundaries and less free neighbors than the others, so we discard minimum hub cover $\{3, 6, 7\}$. Therefore, using our heuristics, the best minimum hub cover and ordering is: $(3, 5, 8)$.

Regarding the use of the data graph, the previous heuristics take only the query graph into account; however, there may be some scenarios in which the data graph helps us to prune our search. For instance, in our motivating example, graphlet 6 has only two possible solutions: graphlets $b$ and $j$ in the data graph. Therefore, to avoid the exploration of a large number of solutions, it is mandatory to discover as soon as possible that graphlet 6 has only $b$ and $j$ as solutions. In these cases, it is possible to store a number of metadata statistics about the data graph to help us prune the search space.
\section{Graph matching}
\label{sec:graph-matching}

\begin{figure}
\begin{listing}[99]
$findSolutions$
input\tab[2]$graph{:} Graph$; $query{:} Graph$; $V{:}$ List of
\tab[6]$Vertex$; $i{:} Integer$; $current{:} Solution$
output\tab $O{:}$ Set of $Solution$
variables $q{:} Graphlet$; $P, R{:}$ Set of $Solution$

// If $i$ is at the end of the order
if $|V| = i$ \label{line:find-solutions:initial-if}
 $O$ := $\{current\}$ \label{line:find-solutions:add-current-solution}
else
 // Substitute the current graphlet using the
 // current solution
 $q$ := $getGraphlet(query, V[i])$ \label{line:find-solutions:get-graphlet}
 $q$ := $substitute(g, current)$ \label{line:find-solutions:substitute}
 // Compute partial solutions by unification
 $P$ := $unify(graph, q)$ \label{line:find-solutions:unify}
 // If the graphlet does not unify
 if $P = undefined$ \label{line:find-solutions:not-unify-1}
  $O$ := $undefined$ \label{line:find-solutions:not-unify-2}
 // If the graphlet unifies but it is ground
 else if $P = \emptyset$ \label{line:find-solutions:empty-unify-1}
  $O$ := $findSolutions(graph, query, V, i + 1, current)$ \label{line:find-solutions:empty-unify-2}
 else
  $O$ := $\emptyset$
  // Iterate over the valid partial solutions
  for each $p{:} Solution \mid p \in P \cdot isValid(p, current)$ do \label{line:find-solutions:valid-iteration}
   $R$ := $findSolutions(graph, query, V, i + 1, p \cup current)$ \label{line:find-solutions:recursive-call}
   // If the recursive solutions unify
   if $R \neq undefined$
    $O$ := $O \cup R$ \label{line:find-solutions:add-partial}
   end if
  end for
  // If the graphlet does not unify
  if $O = \emptyset$ \label{line:find-solutions:no-partial-1}
   $O$ := $undefined$ \label{line:find-solutions:no-partial-2}
  end if
 end if
end if
\end{listing}
    \caption{Algorithm to find the solutions of a query over a data graph.}
    \label{fig:find-solutions}
\end{figure}

Our proposal takes a data graph, a query graph, and an ordering of query vertices to be processed as input. It computes all of the subgraphs by matching the query over the data and, finally, it outputs a number of solutions, each of which comprises the data vertices that match the query vertices. Figure~\ref{fig:find-solutions} shows the algorithm to compute these solutions. Note that, for the sake of brevity, we present the recursive algorithm, whose initial call is the following: $findSolutions(graph, query, V, 0, \emptyset)$.

In our proposal, the recursion ends when we reach the end of the ordering, i.e., $|V| = i$, in which we add the current solution to the final set (see lines~\ref{line:find-solutions:initial-if}--~\ref{line:find-solutions:add-current-solution}). For the rest of the cases, we first substitute the current graphlet of the query graph using the current solution (see lines~\ref{line:find-solutions:get-graphlet}--~\ref{line:find-solutions:substitute}), the intuition behind this is that the current solution may comprise some ground query vertices, which can be replaced in the graphlet by the actual data to narrow the search. With the substituted graphlet, we try to find all of the possible combinations of data vertices for the query vertices (see line~\ref{line:find-solutions:unify}), which result in a number of partial solutions. We perform this by unifying the substituted graphlet with the graphlets in the data graph. Note that a data graphlet $g_d = {<}v_d, N_d, B_d{>}$ unifies with a query graphlet $g_q = {<}v_q, N_q, B_q{>}$ if $|N_q| \leq |N_d|$, $|B_q| \leq |B_d|$, and there exists a solution $s$ such that $s(v_q) = v_d$, $s(N_q) \subseteq N_d$, and $s(B_q) \subseteq B_d$. This unification process depends on the technology used to implement it; additional details are presented in Section~\ref{sec:implementation}.

After the computation of these partial solutions, we have to check if the query has unified with the data. In the case that it does not unify, we output $undefined$ as the final solution, since this mean that the query has no solution (see lines~\ref{line:find-solutions:not-unify-1}--~\ref{line:find-solutions:not-unify-2}). If it unifies and the set of partial solutions is empty, this entails that the query is ground (see lines~\ref{line:find-solutions:empty-unify-1}--~\ref{line:find-solutions:empty-unify-2}), i.e., all of the vertices of the substituted query graphlet belong to the data graph. So we continue with the following vertex in the ordering by performing a recursive call.

Another possibility is that the partial solutions are not empty, in which we iterate over the set of partial solutions that are valid (see line~\ref{line:find-solutions:valid-iteration}). A partial solution $p$ is valid with respect to a current solution $current$ if the query vertices are not repeated ($keys(p) \cap keys(current) = \emptyset$), and the data vertices are not repeated ($values(p) \cap values(current) = \emptyset$). The intuition behind this is that a query and/or data vertex must appear only once in the solution that results from combining $p$ and $current$. Therefore, for each valid partial solution, we continue with the following vertex in the ordering, and the combination of $p$ and $current$ as the next current solution (see line~\ref{line:find-solutions:recursive-call}). If the resulting solutions are not $undefined$, we add them to the output solutions (see line~\ref{line:find-solutions:add-partial}). Finally, if we do not add any solution to the output set, this means that we have found no solution for this query, so we return $undefined$ (see lines~\ref{line:find-solutions:no-partial-1}--~\ref{line:find-solutions:no-partial-2}).

\begin{example}
\label{ex:graph-matching}
To illustrate how our proposal works, we use our motivating example (see Section~\ref{sec:preliminaries}). We also use the following minimum hub cover and ordering: $(3, 6, 7)$; so $findSolutions(graph, query, (3, 6, 7), 0, \emptyset)$ is the first call to our algorithm, in which $graph$ and $query$ comprise all of the data and query graphlets that we presented above. Since $i = 0 \neq |V| = 3$, we retrieve the graphlet of vertex $3$ and substitute it using the current solution, which is empty, so the substituted graphlet is exactly the same.

In the next step, we unify graphlet $3$ with the data graphlets; note that graphlet $3$ comprises 4 neighbors and 2 boundaries, so it is necessary to unify with data graphlets with, at least, 4 neighbors and 2 boundaries. In this case, it can be unified with $b$, $j$, $k$, and $o$ graphlets; however, graphlet $3$ cannot be unified with graphlet $b$ since both boundaries of $b$ comprises vertex $j$, which is not supported by the boundaries of graphlet $3$. For the rest of the graphlets, we compute a number of partial solutions and we iterate over them. All of these partial solutions are valid since the current solution is empty. One of these partial solutions is the following: $p_{01} = \{3 = k, 1 = a, 2 = j, 4 = l, 6 = m\}$; then, we perform a recursive call: $findSolutions(graph, query, (3, 6, 7), 1, p_{01})$.

In this recursive call $i = 1 \neq |V| = 3$, so we retrieve the graphlet of vertex $6$ and substitute it using the current solution, which results in ${<}m, \{k, l, 5, 8\}, \{\{k, l\}\}{>}$ graphlet. We unify it with the data graph and retrieve two valid partial solutions: $p_{11} = \{5 = l, 8 = n\}$ and $p_{12} = \{5 = n, 8 = l\}$. We iterate over them and perform recursive calls; unfortunately, both partial solutions do not unify with the data graph, i.e., it does not exist a data graphlet with ${l, n}$ neighbors and zero boundaries. Therefore, we return $undefined$ as the solution of this recursive call.

Another valid partial solution is $p_{02} = \{3 = k, 1 = l, 2 = m, 4 = a, 6 = j\}$. We perform a recursive call in which $i = 1$, and we retrieve and substitute graphlet $6$ using $p$, which results in ${<}j, \{k, a, 5, 8\}, \{\{k, a\}\}{>}$ graphlet. We unify this graphlet with the data graph and retrieve several valid partial solutions, e.g., $p_{13} = \{5 = c, 8 = e\}$. In the next step, we call $findSolutions(graph, query, (3, 6, 7), 2, p_{02} \cup p_{13})$ and substitute graphlet $7$ with the current solution. We obtain the following graphlet: ${<}7, \{c, e\}, \{\}{>}$, which unifies with the data graph and produces only one partial solution: $p_{21} = \{7 = d\}$. Finally, we call $findSolutions(graph, query, (3, 6, 7), \\ 3, p_{02} \cup p_{13} \cup p_{21})$, in which $i = 3 \neq |V| = 3$, so we add the following solution to the final set: $\{1 = l, 2 = m, 3 = k, 4 = a, 5 = c, 6 = j, 7 = d, 8 = e\}$.
\end{example} 
\section{Implementation}
\label{sec:implementation}

\begin{figure}
\centering
\begin{minipage}{0.45\hsize}
\begin{listing}
<graph>
 <graphlet vertex=``a''>
  <neighbor>b</neighbor>
  <neighbor>j</neighbor>
  <neighbor>k</neighbor>
  <boundary>
   <vertex>b</vertex>
   <vertex>j</vertex>
  </boundary>
  <boundary>
   <vertex>j</vertex>
   <vertex>k</vertex>
  </boundary>
 </graphlet>
 \dots
</graph>
\end{listing}
\end{minipage}
\quad
\begin{minipage}{0.45\hsize}
\begin{listing}
<graph>
 <graphlet vertex=``1''>
  <neighbor>2</neighbor>
  <neighbor>3</neighbor>
  <boundary>
   <vertex>2</vertex>
   <vertex>3</vertex>
  </boundary>
 </graphlet>
 <graphlet vertex=``2''>
  <neighbor>1</neighbor>
  <neighbor>3</neighbor>
  <boundary>
   <vertex>1</vertex>
   <vertex>3</vertex>
  </boundary>
 </graphlet>
 \dots
</graph>
\end{listing}
\end{minipage}
    \caption{Graph examples in XML.}
    \label{fig:xml-graph-examples}
\end{figure}

We implemented a research prototype of our proposal using Java~1.6, eXist-DB~2.1, and Jena~2.10.1. Furthermore, we used Guava~15.0 to implement auxiliary operations, and JGraph~2.2 to draw graphs in a GUI utility. Every graph is transformed into its graphlets, which are stored as an XML object in eXist-DB. Figure~\ref{fig:xml-graph-examples} presents a part of the XML object of the data and query graphlets in our motivating example. The unification process in our implementation is divided into two steps, namely: XQuery extraction and SPARQL unification.

In the first step, our goal is to select those graphlets in the data graph that can be unified with the single query graphlet. Recall that this query graphlet may comprise also ground data, so we take the query graphlet as input and output a query in XQuery to retrieve the data graphlets. In Example~\ref{ex:graph-matching}, we retrieve all of the data graphlets that can unify with graphlet $3$, which comprises 4 neighbors and 2 boundaries, so we issue the following query:

\begin{listing}

for \$x in doc(``DataGraph.xml''/graph/graphlet)
    where count(\$x/neighbor) $\geq$ 4 and
          count(\$x/boundary) $\geq$ 2
return \{\$x\}

\end{listing}

Furthermore, we also retrieve the data graphlets that can unify with graphlet ${<}m, \{k, l, 5, 8\}, \{\{k, l\}\}{>}$, which also comprises 4 neighbors and 2 boundaries, so we issue the following query:

\begin{listing}

for \$x in doc(``DataGraph.xml''/graph/graphlet)
let \$b := \$x/boundary
    where count(\$x/neighbor) $\geq$ 4 and
          count(\$x/boundary) $\geq$ 2 and
          \$x[@vertex = ``m''] and
          \$b/vertex = ``k'' and
          \$b/vertex = ``l''
return \{\$x\}

\end{listing}

In the second step step, we wish to retrieve all of the possible combinations of data vertices. To perform this, we iterate over the whole set of data graphlets that we have retrieved in the previous set. We transform each of them into a set of RDF triples, e.g., the following triples correspond to the RDF transformation of data graphlet $a$ in our motivating example:

\begin{listing}

a       neighbor    b, j, k ;
        boundary    {\_:}B1, {\_:}B2 .
{\_:}B1    vertex      b, j .
{\_:}B2    vertex      j, k .

\end{listing}

Then, we transform the data graphlet into a SPARQL query that is posed over the RDF data. In Example~\ref{ex:graph-matching}, for query graphlet $3$, we issue the following query:

\begin{listing}

SELECT DISTINCT ?1 ?2 ?3 ?4 ?6
WHERE \{
 ?3     neighbor    ?1, ?2, ?4, ?6 ;
        boundary    {\_:}X1, {\_:}X2 .
 {\_:}X1   vertex      ?1, ?2 .
 {\_:}X2   vertex      ?4, ?6 .
\}

\end{listing}

Note that this query may retrieve the same values for different variables that must be filtered out. Furthermore, for query graphlet ${<}m, \{k, l, 5, 8\}, \{\{k, l\}\}{>}$, we issue the following query:

\begin{listing}

SELECT DISTINCT ?5 ?8
WHERE \{
 m      neighbor    k, l, ?5, ?8 ;
        boundary    {\_:}X1 .
 {\_:}X1   vertex      k, l .
\}

\end{listing}
\section{Evaluation}
\label{sec:evaluation}

In the evaluation of our proposal, our main goal is to experimentally validate that the heuristics that we define in Section~\ref{sec:mhc} actually select the best minimum hub cover and the best ordering, i.e., there exists an optimal ordering of a minimum hub cover that performs better than any other random ordering of the minimum hub covers or the complete set of vertices. To validate this, we ran our implementation using our motivating example.

\begin{table}
\centering
\begin{rctabular}{|l|l||l|l|}{Ordering & Explored & Ordering & Explored}
$(3, 5, 8)$ &	189   &    $(6, 7, 1)$    &   297 \\
$(3, 8, 5)$	&   189   &    $(6, 7, 2)$    &   297 \\
$(7, 6, 1)$	&   207   &    $(6, 7, 3)$    &   297 \\
$(7, 6, 2)$	&   207   &    $(6, 1, 7)$    &   425 \\
$(7, 6, 3)$	&   207   &    $(6, 2, 7)$    &   425 \\
$(8, 5, 3)$	&   211   &    $(6, 3, 7)$    &   425 \\
$(5, 8, 3)$	&   211   &    $(3, 7, 6)$    &   1081 \\
$(1, 6, 7)$	&   233   &    $(7, 3, 6)$    &   1171 \\
$(2, 6, 7)$	&   233   &    $(1, 7, 6)$    &   2029 \\
$(3, 6, 7)$	&   245   &    $(2, 7, 6)$    &   2029 \\
$(8, 3, 5)$	&   279   &    $(7, 1, 6)$    &   2131 \\
$(5, 3, 8)$	&   279   &    $(7, 2, 6)$    &   2131
\end{rctabular}
\caption{Orderings of the minimum hub covers and the solutions explored.}
\label{tab:ordering-explored}
\end{table}

In our motivating example, our proposal outputs 24 solutions and Table~\ref{tab:ordering-explored} shows the number of solutions explored when using different minimum hub covers and orderings. The best orderings are $(3, 5, 8)$ and $(3, 8, 5)$, each of which explores 189 solutions, which is what we predicted using our heuristics.

In addition to the minimum hub covers, we computed the number of solutions explored when processing all of the query graphlets using all possible orderings. The best ordering among these is $(3, 5, 8, 6, 1, 2, 4, 7)$, in which our proposal explores 309 solutions. Note that the initial vertices correspond to the best ordering of the minimum hub cover, which entails that we have already computed all of the solutions after processing these three vertices. Therefore, we are performing an extra computation that it is not mandatory. On the contrary, the worst ordering is $(5, 2, 7, 1, 8, 4, 6, 3)$, which explores 8815 solutions.

These results experimentally validate that it is better to use a minimum hub cover to process a query graph instead of using any random ordering of these vertices.

\section{Conclusions}
\label{sec:conclusions}

In this paper, we present a new vertex-at-a-time proposal to match a query graph over a data graph. Our proposal is based on the concept of graphlets, each of which comprises a vertex of a graph, all of the immediate neighbors of that vertex, and all of the edges that relate those neighbors. Furthermore, our proposal use minimum hub covers, each of which comprises a subset of vertices in a query graph that account for all of the edges in that query graph. The main challenge of using minimum hub covers is to select the best one among all possible minimum hub covers for a given query graph and the best ordering to be processed. We have discussed three different aspects that must be taken into account, i.e., the connection of the vertices in the minimum hub cover, the selectivity of the vertices and the use of the data graph. As a product of this discussion, we have devised a number of heuristics to select the best minimum hub cover and the best ordering.

We have presented the algorithms to implement our proposal, an implementation of these algorithms that is based on XQuery and RDF, and some evaluation results of this implementation. Our main conclusion is that we have experimentally validated that using our heuristics we are able to select the best minimum hub cover and the best ordering. Furthermore, our results experimentally validate that it is better to use a minimum hub cover to process a query graph instead of using any random ordering of these vertices.

\end{document}